\documentclass{epl}

\title{Contact angle measurements on superhydrophobic Carbon Nanotube Forests :
effect of fluid pressure}
\author{Catherine Journet\inst{1} \and S\'ebastien Moulinet\inst{2} \and Christophe Ybert\inst{1}
Stephen T. Purcell\inst{1} \and Lyd\'eric Bocquet\inst{1}\footnote[1]{Author for correspondance:
lbocquet@lpmcn.univ-lyon1.fr}}
\institute{
  \inst{1} Laboratoire de Physique de la Mati\`ere Condens\'ee et Nanostructures, 
UMR CNRS and  Universit\'e de Lyon,  43 Boulevard du 11 Novembre,
69622 Villeurbanne France\\
\inst{2}  Laboratoire de Physique Statistique, ENS, 24 rue Lhomond, 75231 Paris, France
}

\begin{document}

\maketitle

\begin{abstract}
In this paper the effect of pressure on the contact angle of a water drop on superhydrophobic
Carbon Nanotube (CNT) forests is studied. 
Superhydrophobic CNT forests are obtained from 
a new and simple functionalization strategy, based on the gold-thiol affinity.
Using a specifically devised experimental setup, we then show that these surfaces are able to withstand high excess pressures (larger than $10 {\rm kPa}$) without transiting toward a roughness-invaded state, therefore preserving their low adhesion properties. 
Together with the relatively low technical cost of the process, this robustness versus pressure makes such surfaces very appealing for practical integration into microfluidic systems. 
\end{abstract}

Recent advances in nanomaterials science has made increasingly possible the design of surfaces
with specific and tunable properties. Such surfaces have a broad range of applications 
from microfluidics to biosensors and biomimetics. In this context the recently developped 
fabrication of surfaces covered with vertically aligned and densely spaced carbon nanotubes
(CNT forests) is a promising method of obtaining composite surfaces modulated at a
sub-micron scale, while taking advantage of the unique CNT properties \cite{Lau2003}.
We present here a versatile functionalization strategy for CNT forests using thiol molecules
and apply this strategy to achieve robust superhydrophobic surfaces.

A superhydrophobic character is common to 
many plants, such as lotus leaves, which have developped leaves with a 
superhydrophobic surface as the basis of a self-cleaning mechanism~:
water drops completely roll off, carrying undesirable
particles \cite{Neinhuis}. This self-cleaning or lotus effect is caused by both
the hierarchical roughness of the leaf surface, which is composed
of micrometer sized papillae having nanometer-sized branch-like
protrusions, and the intrinsic hydrophobicity of a
surface layer of epicuticular wax covering these papillae. 
The roughness {amplifies} the natural non-wetting character of the surface,
leading to very large contact angles - close to $180^{\circ}$ - 
for a liquid drop on  the surface. 
This natural ``superhydrophobicity'' has recently gained much attention,
in particular in the group of D. Qu\'er\'e \cite{Neinhuis,Onda,quere2002,lafuma2003}, and has inspired various attempts to mime it using micro-machined surfaces
\cite{quere2002,lafuma2003,Lau2003,Fan2004}.
\par
Beyond the existence of large contact angles, very low adhesion,
characterized by a small contact angle hysteresis,  is a prerequisite for many 
applications in microfluidics. This is
achieved only in the case where the liquid drop remains at the top of the roughness, known
as Cassie state or Fakir effect \cite{quere2002}. Penetration will
occur above an intrusion pressure, $P_{int}$, which can be estimated
in terms of the liquid-vapor surface tension, $\gamma_{LV}$, the bare contact angle $\theta_0$
of the liquid on the (flat) solid and the roughness characteristics
\cite{cottin2003}. Basically, for a roughness characterized by a distance 
$d$ between two adjacent peaks, the order of magnitude 
of this intrusion pressure is given by $P_{int}\sim 2\gamma_{LV} \cos\theta_0/d$, with corrections
associated with the finite height of the roughness \cite{cottin2003}. An intrusion  pressure of one bar is typically attained for roughness with inter-peak distance $d$ of the order of the micron. This simple argument indicates that robust superhydrophobicity, {\it i.e.} the resistance to 
surface penetration at high pressures, requires roughnesses with small lateral scales ($d$) and high aspect ratios. Such biomimetic surfaces are difficult to obtain using standard microlithography, while more involved techniques able to developp paterns below the micron size are prohibitive in terms of technological  cost, and limited in terms of their ability to provide large surfaces.
\par
In this work, we start from an alternative route, recently explored by Lau {\it et al.}\cite{Lau2003}, using CNT forests. The natural roughness of these surfaces make them suitable to generate a superhydrophobic behavior. However, without further treatment, these surfaces are 'hydrophilic' in the  sense that a water drop is not stable and can completely impregnate the surface \cite{Lau2003}. A further functionalization is thus required to obtain superhydrophobicity.  
We propose an original and versatile functionalization of these composite surfaces using thiol molecules, both in the liquid and vapor phase, opening a wide spectrum of potential applications of the CNT forests using various thiol-modified chemicals. In this article this strategy is used to induce a non wetting character of the surface and the functionalized CNT forests consequently exhibit a superhydrophobic, low adhesive behaviour whose robustness against liquid excess pressure is probed by specific contact angle measurements.
\par
The CNT forests are obtained using Plasma Enhanced Chemical Vapor Deposition (PECVD).
A planar silicon substrate is covered with a SiO$_{2}$ layer and a
thin film of Ni catalyst. The substrate is then introduced into a
furnace heated to 750$^{\circ}$C under 20 torr of hydrogen
during 15 min. This step allows the formation of Ni
catalyst islands on the substrate through the sintering of the
thin Ni film \cite{Chhowalla2001}. Then the furnace is pumped down
and ammonia is introduced at 3.5 torr, the temperature being kept
to 750$^{\circ}$C. A dc discharge between the cathode-sample and
the anode is initiated and increased to 600 V while acetylene
(C$_{2}$H$_{2}$) is introduced within a few seconds using
a separate mass flow controller. The depositions are carried out
between 5 and 60 minutes in a stable discharge. 

\begin{figure}
\centering
\includegraphics[height=8cm]{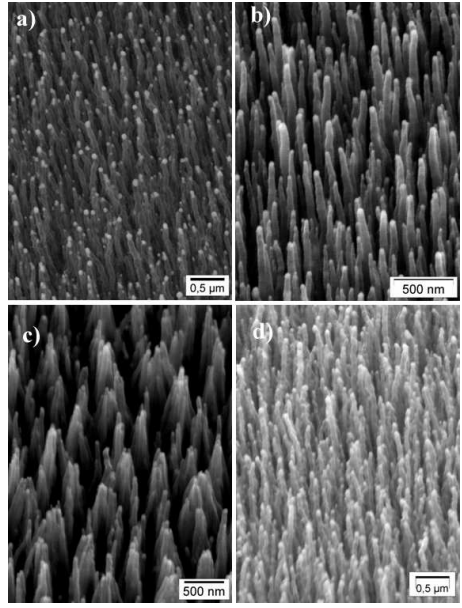}
\caption{{\bf (a)} SEM picture of a forest of well-aligned and straight
carbon nanotubes (5 min deposition time, see text); {\bf (b)} SEM picture
of the CNT forest after deposition of a $5$ nm thick layer of gold,
showing no degradation;
{\bf (c)-(d)} SEM pictures of CNT forests 
after functionalization with thiols : (c) in liquid phase; (d)  
in vapor phase.
\label{fig:1}
}
\end{figure}

Figure \ref{fig:1}(a)-(b) shows typical CNT forests grown through this process,
characterised by a typical nanotube diameter between 50 and 100 nm, a nanotube inter-distance
100-250 nm, and micrometric lengths.
%
\par
The first step of the functionalization consists in covering the
surfaces of CNT forests with a thin gold layer (5 nm) by sputtering, see fig. \ref{fig:1}(b), which forms the anchoring layer for the thiol molecules. Two methods for the subsequent step of attaching the thiols have been tested succesfully. In the first method, the whole CNT forest surface is left during 6 hours in a millimolar solution of alkanethiols (1-octadecanethiol), in absolute ethanol.
%
%
As shown in fig. \ref{fig:1}(c), this functionalization in a liquid phase results in bundles of carbon nanotubes. A similar bundling has been reported in a different context by recent studies \cite{Lau2003,Fan2004,Bico2004,Elasticity}. In the present case, the bundling occurs
during evaporative drying of ethanol and is the result of the competition between attractive 
capillary forces and the elasticity of the carbon nanotubes, along the same lines as discussed
by Bico {\it et al.} for the coalescence of wet hair \cite{Bico2004}.
In the present geometry, the typical bundle size can be estimated as follows.
The elastic free energy of a deformed carbon nanotube 
(with radius $R$ and length $L$) is typically given as
${\cal E}\sim E I/{\cal R}^2 L$, with $E$ the Young's modulus, $I\sim R^4$ the moment of inertia
and ${\cal R}$ is the radius of curvature of the deformation \cite{Landau}. 
In a bundle with lateral size $\xi$, the latter can be estimated to be
${\cal R}^{-1} \sim \xi/L^2$. On the other hand, the capillary energy 
scales as $\gamma_{LV} R L$, with $\gamma_{LV}$ the liquid-vapour
surface tension of ethanol. Energy balance then gives
\begin{equation}
E R^4 {\xi^2 \over L^4} L \sim \gamma_{LV} R L
\end{equation}
leading to a bundle lateral length scale of order 
\begin{equation}
\xi\sim\sqrt{\gamma_{LV}L^4\over {ER^3}}
\label{xi}
\end{equation}
Using for our CNT $E\sim 30\,$GPa \cite{YoungNTC}, $2R\sim 75\,$nm, $L\sim 2\,\mu$m, $\gamma_{LV}\simeq 22\, $mJ.m$^{-2}$, one gets $\xi$ of the order of a few tenths of a micron, in agreement with the observed bundle size as shown in fig. \ref{fig:1}(c). 
A more exhaustive study of the variation of $\xi$ with $L$ or $R$ would be needed
to fully assess the validity of eq. (\ref{xi}). We however leave this specific point for further work.
\par
\begin{figure}
\centering
\includegraphics[height=4cm]{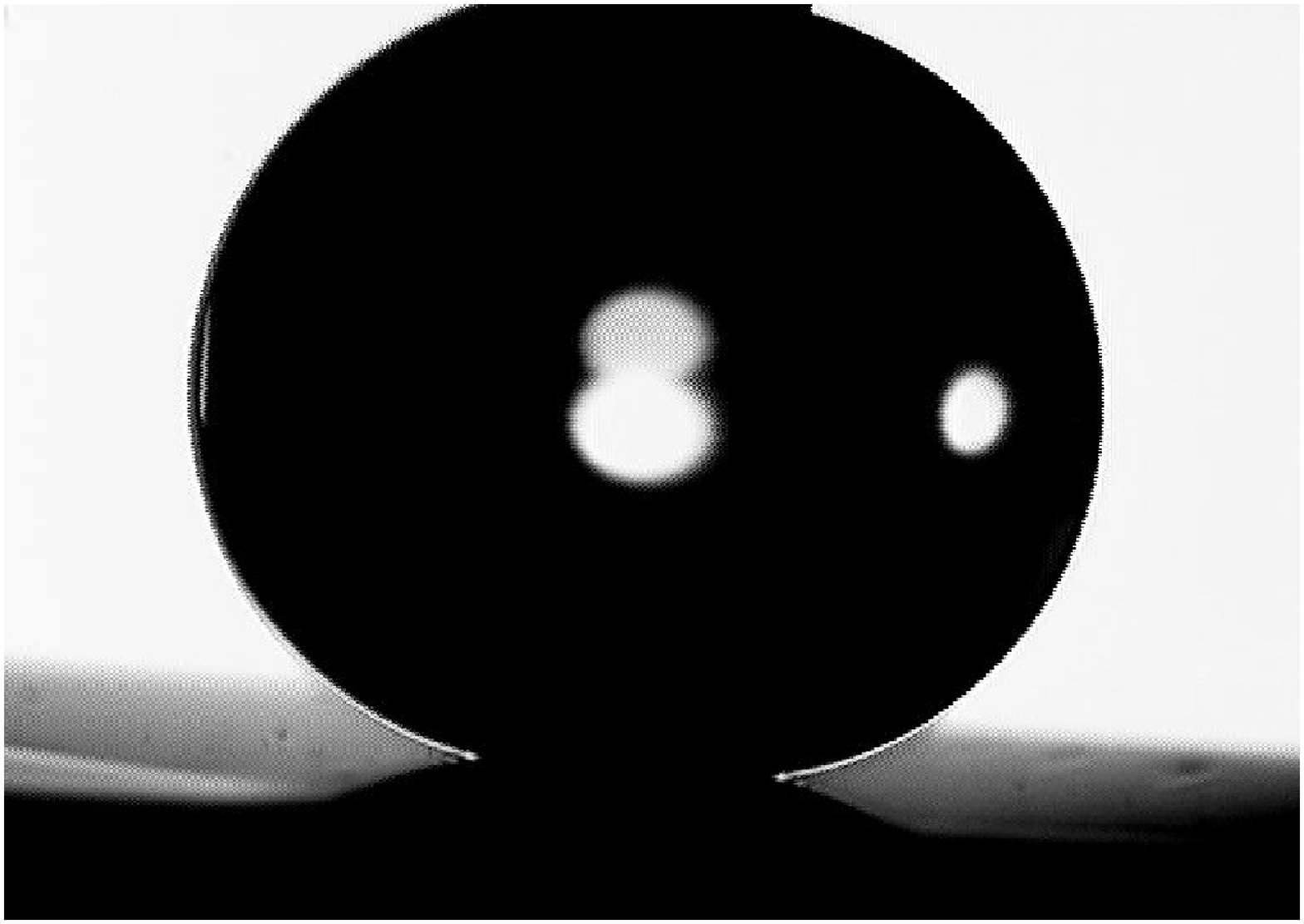}
\includegraphics[height=4cm,width=6cm]{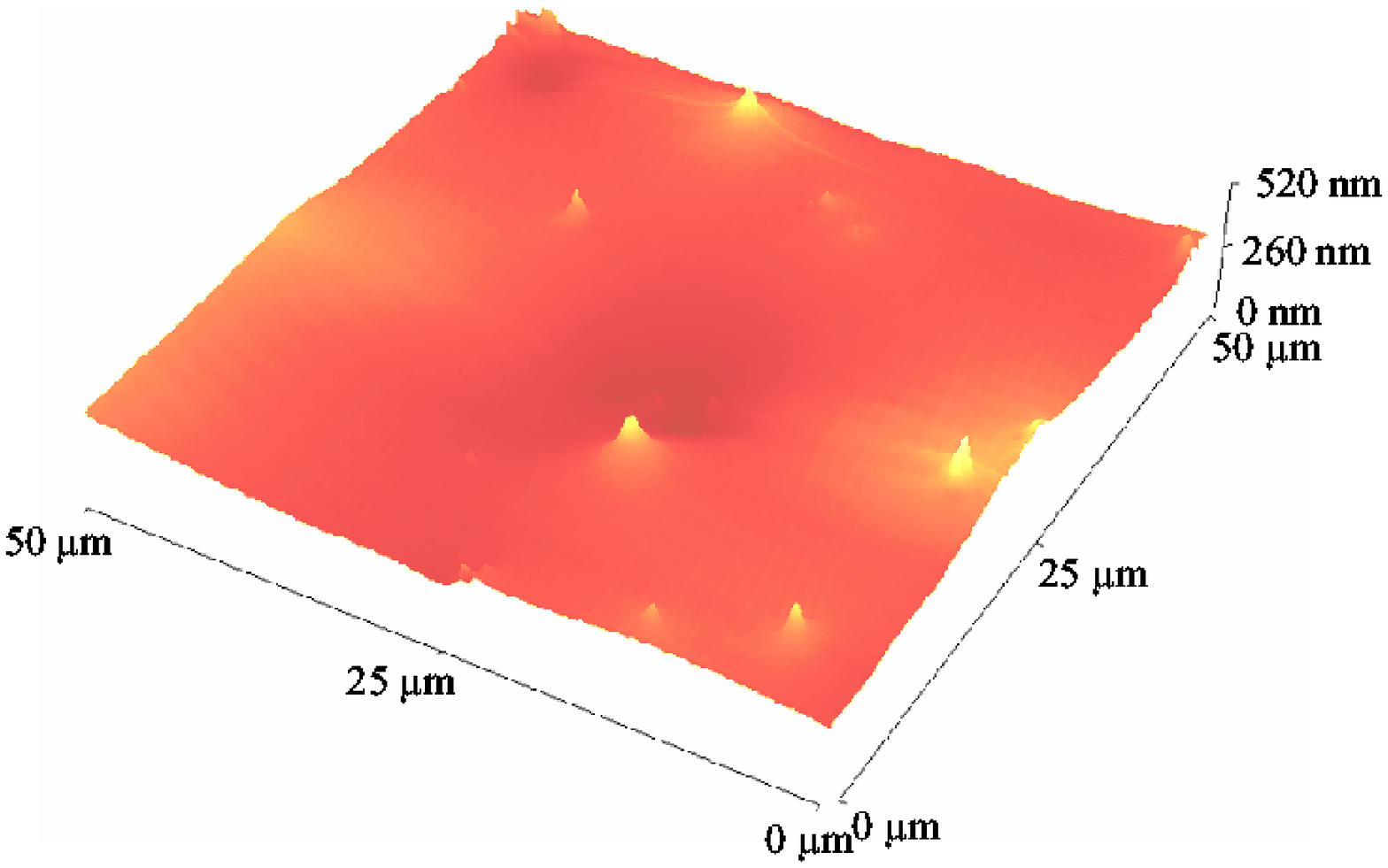}
\caption{ (a) Water drop (of millimetric size) on a nanotube forest, after functionnalisation with thiols,
exhibiting a superhydrophobic character. The drop is held at the end of a seringe to avoid
its immediately running off the surface.
(b) Air-water interface
supported by the tips on the nanotubes, imaged by contact mode AFM in immersion. The light peaks are tips of
nanotube bundles. The radius of the (hydrophilic $SiN_3$) tip apex is about $50\,\mathrm{nm}$
and the exerted force by the tip on the surface is $2\,\mathrm{nN}$. 
\label{fig:3}}
\end{figure}

Despite the bundling of the nanotubes, this functionalized surface does exhibit superhydrophobic properties, as demonstrated in fig. \ref{fig:3} which shows a liquid drop exhibiting a very large contact angle on the surface, $\theta=163^{\circ} \pm 3^{\circ}$ (measured from a direct image analysis of the liquid drop profile from the side).
However, to maintain the nearly ideal geometrical structure of the CNT forest, we have developed
a second method of attaching the thiols, which avoids the bundling. This is based on the functionalization of the surface using thiols in the gas phase. The gold-covered CNT forest
is put in a sealed container together with a reservoir of pure alkanethiols (1-hexadecanethiol, liquid at room temperature). A primary vaccum ($\sim 2\,$torr) is then achieved and, after
the container is isolated from the pump, the partial vapor pressure of volatile thiols is allowed to rise
toward its equilibrium during 6 hours. This functionalization avoids bundling, as shown in fig. 
\ref{fig:1}(d), while preserving the superhydrophobicity. The contact angle is $\theta=164 \pm 2^{\circ}$, equivalent to the one previously obtained. 
These measured values of contact angles on the CNT surfaces are compatible with estimates from the Cassie theory, computed on the basis of the fraction of real solid-liquid contact area, $\phi_S$. 
The latter is estimated from SEM pictures to about $\phi_S\sim0.1$. Using the value of the contact angle on the bare thiol surface (receeding
$\theta_\mathrm{rec}=100^{\circ}$, advancing $\theta_\mathrm{adv}=113^{\circ}$), the Cassie relationship $\cos \theta=\phi_S
\cos \theta_\mathrm{flat} -(1-\phi_S)$ \cite{quere2002,Cassie} leads to the estimate $\theta\simeq 156^{\circ}$--$160^{\circ}$, in fair agreement with the measured value of $164 \pm 2^{\circ}$.
\par
Let us now focus more closely on the wetting properties.  The values
of the contact angle reported above already indicates a superhydrophobic 
character of these surfaces. However, more important is the fact that
{\it no contact angle hysteresis} is measured within experimental errors for a water droplet 
on the CNT forest (see fig. \ref{fig:3}(a)),
indicating that water has not impregnated the CNT forest (Cassie state). 
This is confirmed qualitatively by imaging the liquid-CNT surface using a contact mode AFM
in liquid phase, 
which shows that the air-water interface sits only on the very few highest nanotubes or
bundles, as shown
in fig. \ref{fig:3}(b).
\par
Keeping in mind potential applications, we now demonstrate the robustness of the
superhydrophobicity of the surface. For this, we have developped a new 
experimental procedure to measure the dependance of the contact angle on the CNT forest as a function of the applied excess pressure. Our experimental setup is a modified version of the one proposed by Lafuma and Qu\'er\'e \cite{lafuma2003}, in which a droplet is compressed 
between two surfaces. However in order to reach excess pressure of the order of 
$10\,\mathrm{kPa}$, the gap between the surfaces must be as thin as
$10\,\mathrm{\mu m}$, preventing measures of the contact angles from an imaging of the droplet from the side of the gap. We have thus developed a technique in which the
imaging is performed through one surface.
\par
\begin{figure}[h!]
\begin{center}
\includegraphics[width=7cm]{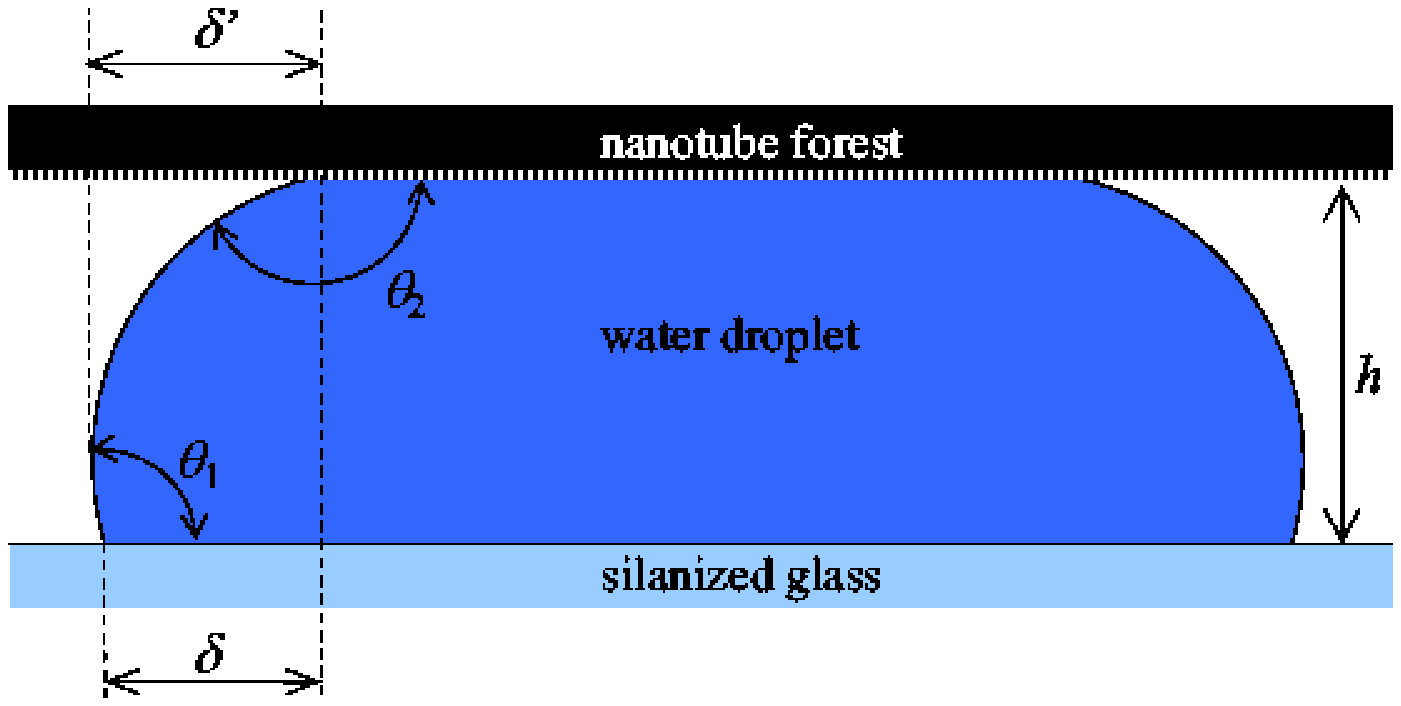}
\includegraphics[width=5cm]{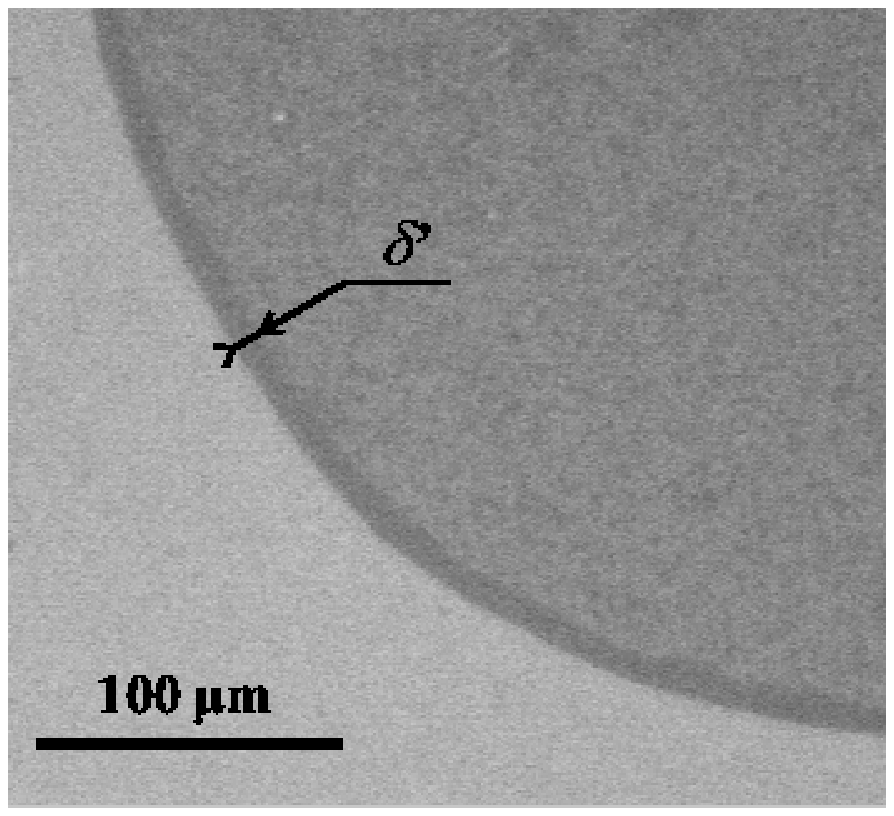}
\end{center}
\caption{(a) schematic configuration of a droplet inside the experimental cell. The projected
widths $\delta$ and $\delta^{\prime}$ are respectively defined as the distance between 
the two contact lines, and the distance between the contact line on the CNT forest and the 
most external point of the droplet profile; (b)
Image of a droplet inside a cell with a 19~$\mu$m wide gap.
The light part inside the drop is in contact with the nanotube
forest. The dark edge is the image of the free surface, with width
$\delta'$. The image is captured by a CCD camera and
the magnifications used allow resolutions of  0.78 and
$0.52\,\mathrm{\mu m}$. } 
\label{fig:4}
\end{figure}

A droplet is first captured inside a cell composed of the CNT surface under study
and a transparent reference surface, separated by a fixed gap (see fig. \ref{fig:4}(a)). Here, the reference surface is a silanized (octadecyltriclorosilane) plate of glass, characterized by a
receeding angle $\theta_1=102\pm2^{\circ}$ (measured by imaging a millimetric drop from side). 
The distance between the glass reference plate and the CNT forest is imposed
by spacers of different heights and materials: Teflon (100,
$50\,\mathrm{\mu m}$), Kapton ($20\,\mathrm{\mu m}$) and Mylar (25,
$9\,\mathrm{\mu m}$). Experimentally, deionized water is sprayed on
the silanized glass before the cell is closed with the nanotube
substrate.
While the width of the gap is imposed by the spacers, measurement of the 
receeding contact angle is obtained by letting the droplet evaporate (we have checked that both contact lines on glass and nanotubes forest are receeding before taking pictures).
The droplet is imaged with an inverted metallographic microscope
through the glass plate and the contact angle is simply deduced from a geometrical analysis 
of the drop meniscus, as shown in fig. \ref{fig:4}.
Depending on how the thickness $h$ compares with the microscope depth of field, it 
is convenient to measure one of the projected widths $\delta$ or $\delta^{\prime}$, as defined
in fig. \ref{fig:4}.
The value of the contact angle $\theta_2$ on the CNT surface is deduced from $\delta$ or $\delta^{\prime}$ using the geometrical expressions (assuming a circular shape of
the meniscus, which is valid in the limit where $h$ is much smaller than the
radius of the drop)~: $\delta/h={(\sin\theta_2-\sin\theta_1)}/{(\cos\theta_1+\cos\theta_2)}$ and ${\delta'}/{h}={(\sin\theta_2-1)}/{(\cos\theta_1+\cos\theta_2)}$. 
Note that gravity can be neglected here since $h$, at most of the order of
a few tens of microns, is much smaller than the millimetric capillary length.
On the other hand, the pressure inside the droplet is given by Laplace's law, $\Delta
p=\gamma_{LV}/R_\mathrm{curv}$, where $\gamma_{LV}$ is the water surface tension and
$R_\mathrm{curv}$ is the radius of curvature of the droplet profile.
The radius $R_\mathrm{curv.}$ can also be deduced from our images
according to $R_\mathrm{curv}={h}/{(\cos\theta_1+\cos\theta_2)}$.
\par
\begin{figure}[h!]
\includegraphics[width=8cm]{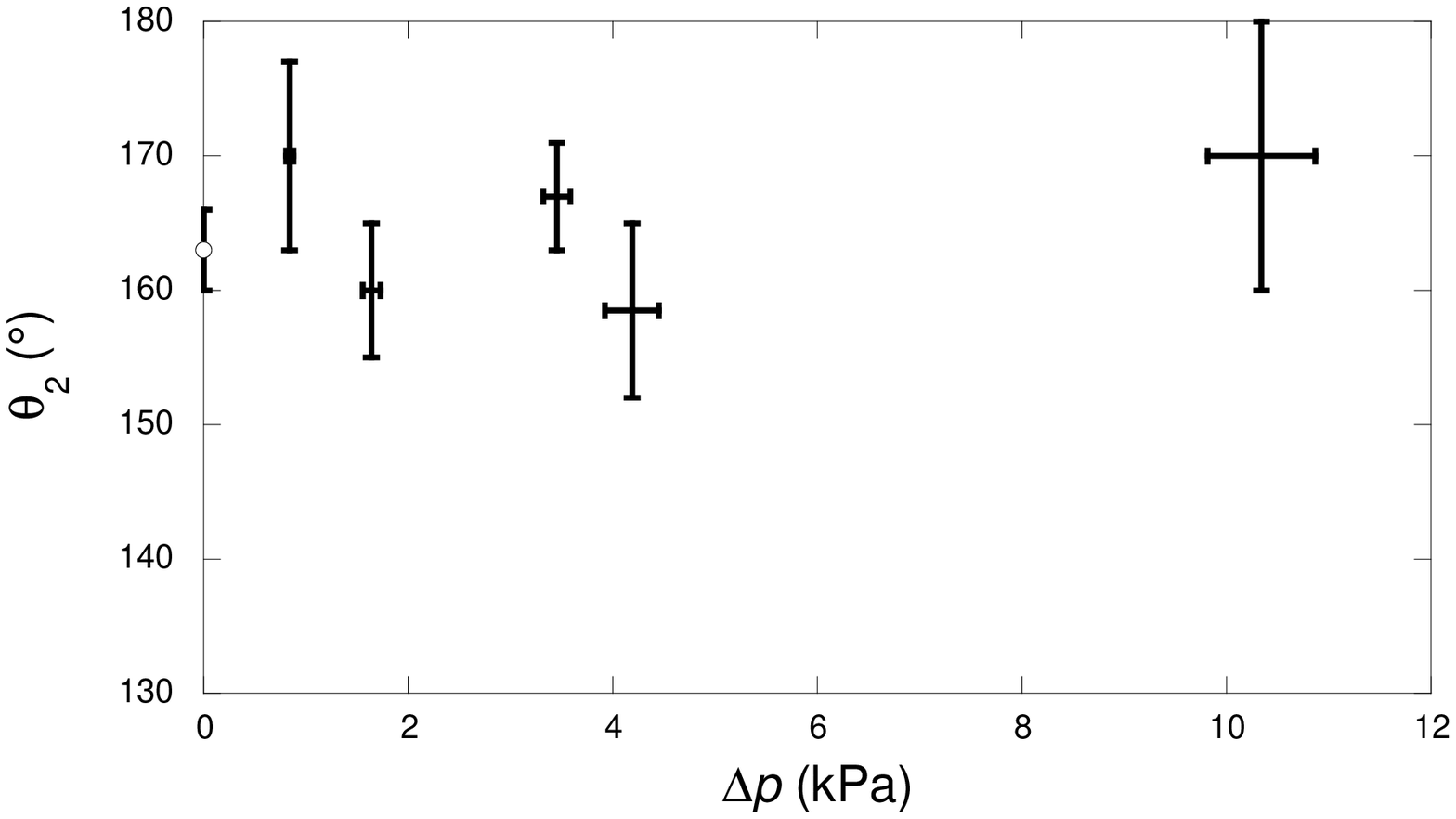}
\caption{Receeding contact angle on the nanotube forest as a function
of the excess pressure. The experimental point at zero pressure is
measured on a millimetric droplet, using a direct imaging of the droplet 
profile, as in fig. \ref{fig:3}(a).} \label{fig:5}
\end{figure}

Figure~\ref{fig:5} shows the receeding contact angle on the
CNT forest as a function of the capillary pressure for $h$ ranging from 8 to $99\,\mathrm{\mu m}$. Error bars take into account the uncertainty on $\delta$ (or $\delta'$) and $\theta_1$. The important point 
is that the
contact angle does not show any significant variation as a function of pressure
up to the highest pressures we could achieve (while measuring contact angles), $\Delta p\simeq10\,\mathrm{kPa}$. 
This robustness originates in the sub-micron scale of NTC forest, which as noted should lead to
intrusion pressure $P_{int}$ of the order of one bar. 
This is in contrast
to surfaces obtained using more conventional lithography, for which $P_{int}$ 
is much lower ($\sim1$kPa), due to
the scale of the roughness. 
As an example, an intrusion pressure of $\sim 0.2$ kPa, with a decrease
of the contact angle and very
large contact angle hysteresis above this value, was obtained on
textured surfaces with 2 micrometers spikes (in height and spacing) in Ref. \cite{lafuma2003}.
This value is to compare with pressures up $10$ kPa reached here without hysteresis, 
nor variation of the contact angle. 
Moreover, while our measurement only yields the receeding contact
angle, this fixes an upper limit for the hysteresis (since the advancing
angle is larger than the receeding one). This indicates
that the hysteresis never exceeds the low value of $15^\circ$, this estimate
being probably far too conservative. One may conjecture that the
absence of hysteresis observed at vanishing excess pressure is
kept up to the largest pressures. This shows that the very low adhesion of 
these surfaces is highly robust and points to potential applications, in particular
in microfluidics. 
\par
The next step towards the aim of using these surfaces in applications is to elaborate 
similar CNT forests in microchannels, in the context of superlubrifying surfaces \cite{cottin2003,Ou}.  Moreover the versatile functionalization proposed here opens the possibility of using the present CNT forest in various contexts, by choosing adequate
chemical grafting among the numerous available thiol-modified chemical compounds.
This opens up new paths towards, for instance, nanotube based biosensors \cite{Func}.
\par

\par
The authors thank Elisabeth Charlaix and Jean-Louis Barrat for many fruitful discussions, 
C. Tassius and M. Monchanin for suggestions and technical support on functionalized CNT surface preparation, P. Joseph and J. Drappier for help in the contact angle measurements.

\end{document}